# Unusual Magnetic Hysteresis and Transition Between Vortex to Double Pole States Arising from Interlayer Coupling in Diamond Shaped Nanostructures

A. Parente[1], H. Navarro[2], N. M. Vargas[2], P. Lapa[2], Ali C. Basaran[2], E. M. González[1,3], C. Redondo[4], R. Morales[5,6], A. Munoz Noval[1,3]*, Ivan K. Schuller[2] and J. L. Vicent[1,3]

1. Departamento de Física de Materiales, Facultad de Física, Universidad Complutense de Madrid, Pl. Ciencias, 1, 28040, Madrid, Spain.
2. Department of Physics and Center for Advanced Nanoscience, University of California San Diego, La Jolla, CA, 92093, USA
3. IMDEA Nanociencia, c/ Faraday, 9, 28049 Madrid, Spain
4. Department of Physical Chemistry, University of the Basque Country UPV/EHU, 48940, Leioa, Spain
5. Department of Physical Chemistry, University of the Basque Country UPV/EHU & BCMaterials, 48940 Leioa, Spain
6. IKERBASQUE. Basque Foundation for Science, 48009 Bilbao, Spain

* corresponding author: alvaro.munoz.noval@ucm.es

## Abstract

*Controlling the magnetic ground states at the nanoscale is a long-standing basic research problem and an important issue in magnetic storage technologies. Here, we designed a nanostructured material that exhibits very unusual hysteresis loops due to a transition between vortex and double pole states. Arrays of 700 nm diamond-shape nanodots consisting of Py(30 nm)/Ru($t_{Ru}$)/Py(30 nm) (Py, permalloy ($Ni_{80}Fe_{20}$)) trilayers were fabricated by interference lithography and e-beam evaporation. We show that varying the Ru interlayer spacer thickness ($t_{Ru}$) governs the interaction between the Py layers. We found this interaction mainly mediated by two mechanisms: magnetostatic interaction that favors antiparallel (antiferromagnetic, AFM) alignment of the Py layers and exchange interaction that oscillates between ferromagnetic (FM) and AFM couplings. For a certain range of Ru thicknesses, FM coupling dominates and forms magnetic vortices in the upper and lower Py layers. For Ru thicknesses at which AFM coupling dominates, the magnetic state in remanence is a double pole structure. Our results showed that the interlayer exchange coupling interaction remains finite even at 4 nm Ru thickness. The magnetic states in remanence, observed by Magnetic Force Microscopy (MFM), are in good agreement with corresponding hysteresis loops obtained by Magneto-Optic Kerr Effect (MOKE) and micromagnetic simulations.*

**Introduction**

The rapid growth of data management and storage caused by the advent of big data requires the information density to increase, motivating the search for new materials and recording systems[1]. To meet challenges beyond von Neumann's architecture, new computing concepts, unusual functional materials, and novel nanofabrication techniques are proposed[2-4]. Furthermore, magnetic recording technologies based on layered magnetic thin films are still widely used and may exhibit unexpected properties. The combination of unique concepts, such as perpendicular magnetic anisotropy[5], giant magnetoresistance[6], oscillating interlayer coupling[7,8], and skyrmion textures[9], may give rise to unexpected emergent results and potential novel applications. Layered magnetic films combined with nanofabrication techniques have been used to develop, for example, new functional nanoparticles for biomedical applications[7-9]. Similarly, magnetic nanodisks with high saturation magnetization and zero remanence are synthesized to improve remote manipulation[10-14]. Alternatives to traditional recording techniques rely on increasing the number of memory states in a single bit[15-17]. However, stabilizing multi-state magnetization states at relevant length and time scales seems not to be possible for practical applications.

On one hand, shape anisotropy caused by size reduction can result in topological spin textures such as magnetic vortices[18-20], skyrmions[21], helical spin structures[22,23], and multi-polar nanomagnets[24-26]. These are mostly characterized and imaged by advanced synchrotron techniques[27]. Ferromagnetic (FM) / antiferromagnetic (AFM) coupling of magnetic vortex states to stabilize the same (FM) or opposite (AFM) chirality in trilayer structures has been recently observed [28-30]. These studies have paved the way to develop new concepts in magnetic memories, for instance, by controlling vortex polarity or chirality[22,31-35]. To obtain and manage such complex magnetic states placing a non-magnetic spacer between two magnetic layers can be used to stabilize competing energy landscapes or discover exciting new properties. For example, the oscillating interlayer exchange coupling (IEC) between magnetic layers separated by a non-magnetic spacer, such as ruthenium[5], is useful to control the magnetic configuration of ferromagnetic multilayers[36,37]. This type of research in continuous layered systems has been pursued for a long time. On the other hand, combining interlayer exchange coupling with lateral confinement is much less explored and may lead to unusual, unexpected results.

In this work, we discovered some very unusual hysteresis loops, unlike any other magnetic system. This was found by a systematic study of the interlayer magnetic exchange interaction in diamond shaped Py/Ru($t_{Ru}$)/Py trilayer nanodots with four-fold symmetry combining magneto-optic Kerr effect (MOKE), magnetic force microscopy (MFM), and micromagnetic simulation. The interaction between interlayer coupling and the boundaries of diamond-shaped nanodots produce interesting magnetic structures which can be visualized[38] using MFM. We observe a crossover from magnetic vortex to magnetic double pole state as a function of the Ru interlayer spacer thickness ($t_{Ru}$), a consequence of the interlayer exchange interaction strength change. Increasing the Py layer separation progressively decreases the interlayer exchange coupling, which tends to align the Py magnetizations parallel. The magnetostatic interaction between Py layers forces the magnetization to align antiparallel and as a consequence, interesting magnetic hysteresis arises unlike any earlier observations. These results show that fine-tuning the inter-layer magnetic interaction provides new degrees of freedom and can be adapted to develop new memory devices based on spin textures.

**Materials and Methods**

The sample structure is summarized in Figure 1. Ruthenium (Ru) spacer ($t_{Ru}$ in nm) between two 30 nm Permalloy layers (Py, ($Ni_{80}Fe_{20}$)) and a 5 nm Pd capping layer were sequentially deposited on Si (100) wafers by electron beam evaporation in a high-vacuum chamber with a base pressure of $10^{-7}$ Torr. For each Ru thickness, Py(30 nm)/Ru($t_{Ru}$)/Py(30 nm) reference continuous thin films and lithographically patterned arrays were simultaneously deposited (Figure 1a). For patterned films, 700 nm 2D square arrays of antidots were initially fabricated by laser interference lithography (LIL)[10]. A photosensitive stack made using a bottom antireflective coating WIDE-8B (from Brewer Science), and negative resist tone TSMR-IN027 (from Ohka) were spin-coated on silicon wafers. The photosensitive layer was then illuminated by the interference pattern of a 325 nm wavelength laser. The regions exposed to the highest light intensity remain on the substrate, while underexposed areas became soluble to the developer AZ 736 MIF, leading to an antidot array in the resist. By e-beam evaporation, an array of dots with a trilayer structure, shown in Figure 1b was prepared. After film deposition, the resist templates were removed by N-methyl pyrrolidone and low-power sonication for about 30 min. Layer thicknesses were controlled in situ by a quartz microbalance and later confirmed by X-ray reflectometry (XRR) for each reference sample of the continuous film (Figure 1c). The profile fitting of the XRR measurements was made with the *Parrat32* code[39], which presents well-defined thicknesses with an interlayer roughness of less than 1.3 nm. The asymptotic dependence of the interface roughness with spacer thickness implies that the interlayer coupling is not affected by dipolar interaction or interface roughness[40]. The successful lift-off of the resist was verified in every sample by optical and scanning electron microscopy (SEM) (Figure 1d).

A commercial Park XE7 AFM-MFM system was employed to study the magnetic textures of the arrays. The probes are commercial Si cantilevers Nanosensors PPP-MFMR, $k =$ 2.8 N/m and $f =$ 75 kHz coated by a Co thin film. Before each measurement, the probes are magnetized by applying 0.8 T along their pyramid axis. The MFM contrast achieved with these probes can resolve the chirality and polarity of the vortex and the double pole spin textures.

Magneto-Optical Kerr Magnetometry measurements were performed in a Durham Magnetooptics NanoMOKE3 system at room temperature. The effect of interdot interaction on the MOKE was addressed by performing magnetization loops with the external field applied parallel and diagonal to the lattice array direction, which corresponds to different interdot distances. In both configurations, the hysteresis loops conserve the same shapes. To ensure the reproducibility of the results between different nanostructures, MOKE measurements were acquired in each sample at least in four different locations.

The micromagnetic simulations were performed using a GPU-accelerated micromagnetic simulation program, MuMax3[41]. The simulations were obtained considering two separated Py disks with saturation magnetization $M_S = 8.6 \cdot 10^5 \, A/m$, exchange stiffness $A_{ex} = 1.3 \cdot 10^{-11} \, J/m$, and 0.5 Landau-Lifshitz damping constant[42]. The mesh cell size is 10.9 × 10.9 nm in the plane of the disk and 15 nm over its thickness. No lateral disk-to-disk interaction has been considered for the calculations. The interlayer exchange interaction has been included using an interlayer exchange parameter that has been varied from $-10^{-3}$ to $+10^{-3} \, J/m^2$.

**Results**

The 700 nm Py nanostructure (60 nm thick) exhibits characteristic magnetic loops of a magnetic vortex state measured by longitudinal MOKE (L-MOKE, Figure 2a). The vortex state[20] has a 40 Oe nucleation field, in good agreement with reported values for this diameter and thickness range[43]. Similar results are obtained for 30 nm thick samples (not shown). The interaction between Py layers of the nanostructures as a function of the spacer thickness was studied using the same MOKE configuration. Figure 2b-i shows the hysteresis loops of the arrays of Py(30)/Ru($t_{Ru}$)/Py(30) trilayer structures. Note that the typical penetration depth of Kerr signal in metallic layers for a 600 nm laser depends on the optical configuration and materials studied[44] and decreases exponentially. For the case of Py, it is about 50 nm[45]. Therefore, the observed signal arises from the upper and lower magnetic layers, with a more significant contribution from the upper layer signal. The 3 μm laser spot size provides enough lateral resolution to avoid inhomogeneities or array defects. The figure 2b ($t_{Ru}$ = 0.3 nm) shows a loop that corresponds to a vortex state with increased coercivity (broadening of the loop "neck"), indicating a nucleation field close to zero. For a fixed diameter, the nucleation field increases with Py [19] thickness. Thus, the reduction of the nucleation field is related to a reduced Py thickness compared to the 60 nm Py sample. In the samples with 0.5, 0.8, 1.5, 2.0, and 3.0 nm Ru spacer, the magnetization (extracted from Kerr rotation) drops before zero in the positive reversing branch for the nonzero applied field. Similarly, the Kerr rotation flips sign when approaching zero applied field for decreasing reversing field. When the external field inverts, these curves present a kink in the direction of decreasing Kerr rotation. It should be noted that these very unusual magnetic hysteresis curves depend on the thickness of the Ru spacer. For the samples with $t_{Ru}$ = 1.0 nm, and 4.0 nm, the loops display a narrow neck at low fields indicative of a vortex state, and there is no crossing of the magnetic reversal branches.

To further understand these unusual hysteresis loops arising from complex spin configurations, MFM images were obtained. Measurements were performed in the samples in remanence after several magnetization cycles. The diamond-like shape of the structures allows observing the helicity of the magnetic vortices due to symmetry break[38,46]. The MFM images in Figure 3 clearly show the formation of magnetic vortices in the single Py and the trilayer with 0.3 nm of Ru nanostructures. In both samples, the chirality and polarity of the vortices seem to be randomly distributed, even after several magnetization cycles. For Ru thicknesses above 0.3 nm, the vortices are not observed in remanence. Interestingly, for thicker Ru spacers, the vortices disappear, and two small poles of the same magnetic polarity are formed in the boundary of the structures. These poles tend preferably to locate in the vertex of the diamond and, in most cases, are antiparallel. The surface magnetization, except for the poles, gives no MFM contrast implying that the magnetization lies in the plane.

Micromagnetic simulations were conducted considering two diamond-shaped layers separated by a distance equal to the corresponding spacer thickness. Two interactions between layers were considered, i) magnetostatic, which favors antiparallel alignment, and ii) interlayer exchange coupling that oscillates in intensity and sign as $E_{ex} \propto J_{ex} \cos(\kappa_F t)$, ($J_{ex}$ the exchange constant, $\kappa_F$ the Fermi wavevector and *t* the spacer thickness). Figure 4 shows the hysteresis loops obtained from the simulations for the sum of two layers and the upper and bottom layers separately. Negative, positive, and zero values for the exchange interaction are also included. For positive exchange parameters larger than $\sim 5 \cdot 10^{-4}$ J/m$^2$, hysteresis loops of the upper and lower layers imply the formation of vortices of the same chirality leading to a parallel alignment of the magnetic moments. This configuration results in a magnetic vortex in remanence (zero applied field). For smaller but still positive exchange interaction below this value results in a sign change of the inter-layer interaction. This occurs because the magnetostatic interaction forces the magnetization of the two layers to align antiparallel. The micromagnetic calculations show that the hysteresis loops of top and bottom layers differ for zero and negative exchange interaction parameters, leading to a hysteresis loop of the whole system with crossing branches in qualitative agreement with MOKE measurements. More remarkably, the magnetic configuration of the upper Py layer in remanence matches the features observed by MFM, showing the formation of two singularities on opposite sides of the structures that result in the double pole state at the upper Py layer.

MFM measurements have allowed identifying the two magnetic configurations that the trilayer nano dots adopt, depending on the spacer thickness. The global picture of the magnetic structure of the double-pole and the magnetic reversal has been achieved by performing the l-MOKE measurements and micromagnetic calculations. From the MOKE, we found a reduction of the vortex nucleation field in the $t_{Ru}$ = 0.3 nm sample, attributed to a reduction of the Py layer thickness. However, for similar dot diameters, an increase of nucleation/annihilation fields is expected when decreasing Py thickness[43]. This opposite trend was observed by P. Vavassori et al. [48]. It was attributed to a competition of an antiferromagnetic coupling caused by magnetostatic dipolar interaction between the magnetic layers and a ferromagnetic coupling due to indirect exchange interaction[49]. Therefore, this evidences the coexistence of, at least, two interactions between the ferromagnetic layers. In the samples with unusual magnetic loops (the $t_{Ru}$ = 0.5, 0.8, 1.5, 2.0, 3.0 nm), during magnetic reversal the magnetization drops before zero, and would be indicative of two effects: i) the nucleation field is significatively decreased, and the magnetic vortex nucleates at very low decreasing external magnetic fields; ii) the zero-magnetization state (i.e., vortex core well centered in the dot) takes place even under small external fields. This can only occur if an internal field compensates for the effect of the external field stabilizing the vortex core. As the external field is decreased to zero, the Kerr signal changes sign. This would be evidence of an AF coupling with the bottom layer that forces the upper layer to align antiparallel[50]. The kink in the increasing negative field direction suggests a decrease in magnetization (Kerr rotation) just after switching the external field polarity. This could be due to an antiparallel alignment of the two Py layers, each one in a vortex state. For this magnetic configuration, there are two vortices in each Py layer with different chirality (antiparallelly aligned), and each one will annihilate at different fields. The magnitude of the kink decreases with increasing Ru thickness and disappears in the $t_{Ru}$ = 3.0 nm sample. In the samples with $t_{Ru}$ = 1.0 nm, and 4.0, there is no crossing of the magnetic reversal branches for these Ru thicknesses. For these Ru thicknesses, the loops display a narrow neck at the low external field region, fingerprint of vortex state.

Remarkably, the behavior of the simulated hysteresis loop of the upper layer coincides qualitatively with experimental data obtained from MOKE measurements. This arises because the penetration depth of the laser (∼50 nm) is of the order of the system's total thickness. Therefore, the Kerr signal has more contribution from the upper layer. The hysteresis loops and magnetic states in remanence obtained by micromagnetic calculations support the experimental data. The use of Ru spacers in magnetic heterostructures is well known to modulate the magnetic interlayer coupling of the magnetic layers[5]. With increasing $t_{Ru}$ the inter-layer interaction oscillates between ferromagnetic and antiferromagnetic. The oscillating exchange contribution to the interlayer coupling progressively weakens for thicker Ru spacers. This behavior in the trilayer nanostructures gives rise to the unusual, crossed cycles, but for 1 and 4 nm Ru spacer thicknesses, that present typical loops of a magnetic vortex state. The crossed hysteresis cycles are correlated with a magnetic state with a double-pole configuration observed by MFM. For samples in this range of Ru thickness, the micromagnetic calculations point to the magnetostatic interlayer interaction overcoming the interlayer exchange coupling, resulting in an antiparallel alignment of both ferromagnetic layers. The detailed analysis of the magnetization map of each layer obtained in the calculations shows that in the double pole state, the magnetization in each layer is forced by the shape anisotropy and aligned to the diagonal of the dot. The upper and bottom layers are aligned antiparallel, and therefore the stray fields close in each of the opposite poles of the dot. The magnetization structure of the trilayer obtained from the calculations, imply that the formation of other complex magnetic configurations such as antivortices is unlikely[51,52].

Table I summarizes the MFM and MOKE results for increasing Ru spacer thickness. A comparison with the sign of the interlayer exchange energy parameter, $J_{ex}$, obtained from the literature, agrees with the proposed model of two competing interactions between the Py layers. These are: i) an interlayer exchange coupling that is FM for certain thicknesses of Ru spacers and decreases in intensity with increasing thickness, ii) a magnetostatic coupling between the two Py layers that favours antiparallel alignment. The coexistence of both interactions implies that the FM/AFM coupling of the layers oscillates in this range of Ru thickness and favours the formation of a vortex or a double-pole state.

It should be noted that the observation of a coherent distribution of magnetic vortices by MFM has been only achieved at $t_{Ru}$= 0.3 nm, but not for the other FM coupled conditions (1.0 and 4.0 nm of Ru spacer). This is probably related to the rapid decrease of FM interlayer exchange coupling with the spacer thickness, which is at least one order of magnitude lower than 0.3 nm Ru thickness[36,47]. For $t_{Ru}$ = 1.0 nm, not all observed diamond shape structures present a vortex state in remanence, and for $t_{Ru}$= 4.0 nm, vortices are observed only occasionally. The oscillatory IEC in FM layers with Ru spacers depends critically on the $J_{ex}$ sign, which is very sensitive to the spacer thickness. For certain values, the $J_{ex}$ can flip in sign by slight variations of $t_{Ru}$. We argue that experimentally, although the nominal thickness of the spacer layer is well defined, slight variations may present, which can cause these isolated vortices.

The interlayer exchange coupling sign and amplitude as a function of the Ru thickness are in good agreement with the spacer diagram by Parkin et al.[36] and Bloemen et al.[38]. Our results show that interlayer exchange coupling depends critically on the Ru thicknesses, particularly for thinner spacers, for which slight variations in $t_{Ru}$ may change the exchange interaction ($J_{ex}$) sign.

## Conclusions

We found a crossover from a magnetic vortex to a double magnetic pole state as a function of Ru thickness in diamond-shaped Py/Ru/Py trilayer nanostructures. The unusual magnetic hysteresis loops can be understood as a consequence of two coupling mechanisms between the Py layers: magnetostatic interaction that favors antiparallel alignment and interlayer exchange coupling that oscillates between ferromagnetic (FM) and antiferromagnetic (AFM) alignment. Fine-tuning the Ru spacer thickness to achieve FM coupling between the Py magnetic layers, results in the formation of magnetic vortices in the upper and lower layers. The largest FM interaction observed corresponds to the thinnest Ru spacer, although even at 4.0 nm Ru thicknesses there are still observable effects due to the FM coupling. Furthermore, double pole magnetic structures are observed when the magnetostatic coupling prevails, and the layers are antiparallelly coupled.

## Acknowledgments

The authors acknowledge Prof. J. M. Alameda and Prof. G. Armelles for fruitful discussion. This work was supported by Comunidad de Madrid and Universidad Complutense de Madrid under the project 2018-T1/IND-10360 granted by "Atracción de Talento" program, the European Union under the H2020 research and innovation Marie Sklodowska-Curie Grant Agreement MAGNAMED 734801 (H2020-MSCA-RISE-2016), the Spanish MINECO grants FIS2016-76058 (AEI/FEDER, UE), PID2019- 104604RB/AEI/ 10.13039/501100011033, PID2021-122980OB-C52 and Basque Government grant IT1491-22. The magnetic measurements and interpretation were funded by the Department of Energy's Office of Basic Energy Science, under grant # DE-FG02-87ER45332. IMDEA Nanociencia acknowledges support from the 'Severo Ochoa' Programme for Centres of Excellence in R&D (MICINN, Grant CEX2020-001039-S). The design of the experiments and the multiple versions of the manuscript were discussed and written with extensive contributions from all the authors.

## References

1 Vedmedenko, E. Y. *et al.* The 2020 magnetism roadmap. *Journal of Physics D: Applied Physics* **53**, 453001 (2020).

2 Grollier, J., Querlioz, D. & Stiles, M. D. Spintronic Nanodevices for Bioinspired Computing. *Proceedings of the IEEE* **104**, 2024 (2016).

3 Liu, C. *et al.* Two-dimensional materials for next-generation computing technologies. *Nature Nanotechnology* **15**, 545 (2020).

4 Wang, K., Bheemarasetty, V., Duan, J., Zhou, S. & Xiao, G. Fundamental physics and applications of skyrmions: A review. *Journal of Magnetism and Magnetic Materials* **563**, 169905 (2022).

5    Parkin, S. S. P., More, N. & Roche, K. P. Oscillations in exchange coupling and magnetoresistance in metallic superlattice structures: Co/Ru, Co/Cr, and Fe/Cr. *Physical Review Letters* **64**, 2304 (1990).

6    Baibich, M. N. *et al.* Giant Magnetoresistance of (001)Fe/(001)Cr Magnetic Superlattices. *Physical Review Letters* **61**, 2472 (1988).

7    Parkin, S. S. P. Systematic variation of the strength and oscillation period of indirect magnetic exchange coupling through the 3d, 4d, and 5d transition metals. *Physical Review Letters* **67**, 3598 (1991).

8    Lapa, P. N. *et al.* Magnetization reversal in Py/Gd heterostructures. *Physical Review B* **96**, 024418 (2017).

9    Fert, A., Reyren, N. & Cros, V. Magnetic skyrmions: advances in physics and potential applications. *Nature Reviews Materials* **2**, 17031 (2017).

10   Mora, B., Perez-Valle, A., Redondo, C., Boyano, M. D. & Morales, R. Cost-Effective Design of High-Magnetic Moment Nanostructures for Biotechnological Applications. *ACS Applied Materials & Interfaces* **10**, 8165 (2018).

11   Dutz, S. & Hergt, R. Magnetic particle hyperthermia—a promising tumour therapy? *Nanotechnology* **25**, 452001 (2014).

12   Lapa, P. N. *et al.* Magnetic vortex nucleation/annihilation in artificial-ferrimagnet microdisks. *Journal of Applied Physics* **122**, 083903 (2017).

13   Zamay, T. N. *et al.* Magnetic Nanodiscs-A New Promising Tool for Microsurgery of Malignant Neoplasms. *Nanomaterials (Basel)* **11**, 1459 (2021).

14   Peixoto, L. *et al.* Magnetic nanostructures for emerging biomedical applications. *Applied Physics Reviews* **7**, 011310 (2020).

15   Shinjo, T., Okuno, T., Hassdorf, R., Shigeto, K. & Ono, T. Magnetic Vortex Core Observation in Circular Dots of Permalloy. *Science* **289**, 930 (2000).

16   Bussmann, K., Prinz, G. A., Cheng, S.-F. & Wang, D. Switching of vertical giant magnetoresistance devices by current through the device. *Applied Physics Letters* **75**, 2476 (1999).

17   Morales, R., Kovylina, M., Schuller, I. K., Labarta, A. & Batlle, X. Antiferromagnetic/ferromagnetic nanostructures for multidigit storage units. *Applied Physics Letters* **104**, 032401 (2014).

18   Guslienko, K. Y., Novosad, V., Otani, Y., Shima, H. & Fukamichi, K. Field evolution of magnetic vortex state in ferromagnetic disks. *Applied Physics Letters* **78**, 3848 (2001).

19   Guslienko, K. Y., Novosad, V., Otani, Y., Shima, H. & Fukamichi, K. Magnetization reversal due to vortex nucleation, displacement, and annihilation in submicron ferromagnetic dot arrays. *Physical Review B* **65**, 024414 (2001).

20   Ky, G. Magnetic vortex state stability, reversal and dynamics in restricted geometries. *Journal of Nanoscience and Nanotechnology* **8**, 2745 (2008).

21   Li, Y. *et al.* An Artificial Skyrmion Platform with Robust Tunability in Synthetic Antiferromagnetic Multilayers. *Advanced Functional Materials* **30**, 1907140, (2020).

22      Vargas, N. M. *et al.* Helical spin structure in iron chains with hybridized boundaries. *Applied Physics Letters* **117**, 213105 (2020).

23      Torres, F., Kiwi, M., Vargas, N. M., Monton, C. & Schuller, I. K. Chiral symmetry and scale invariance breaking in spin chains. *AIP Advances* **10**, 025215 (2020).

24      Escobar, R. A. *et al.* Complex magnetic reversal modes in low-symmetry nanoparticles. *Applied Physics Letters* **104**, 123102 (2014).

25      Escobar, R. A. *et al.* Multi-stability in low-symmetry magnetic nanoparticles. *Journal of Applied Physics* **117**, 223901 (2015).

26      Sinnecker, E. H. d. C. P., García-Martín, J. M., Altbir, D., D'Albuquerque e Castro, J. & Sinnecker, J. P. A Magnetic Force Microscopy Study of Patterned T-Shaped Structures. *Materials* **14**, 1567 (2021).

27      Stebliy, M. E., Kolesnikov, A. G., Ognev, A. V., Samardak, A. S. & Chebotkevich, L. A. Manipulation of magnetic vortex parameters in disk-on-disk nanostructures with various geometry. *Beilstein J Nanotechnol* **6**, 697 (2015).

28      Wurft, T. *et al.* Evolution of magnetic vortex formation in micron-sized disks. *Applied Physics Letters* **115**, 132407 (2019).

29      Wintz, S. *et al.* Direct observation of antiferromagnetically oriented spin vortex states in magnetic multilayer elements. *Applied Physics Letters* **98**, 232511 (2011).

30      Buchanan, K. S. *et al.* Magnetic remanent states and magnetization reversal in patterned trilayer nanodots. *Physical Review B* **72**, 134415 (2005).

31      Van Waeyenberge, B. *et al.* Magnetic vortex core reversal by excitation with short bursts of an alternating field. *Nature* **444**, 461 (2006).

32      Yamada, K. *et al.* Electrical switching of the vortex core in a magnetic disk. *Nature Materials* **6**, 270 (2007).

33      Pigeau, B. *et al.* A frequency-controlled magnetic vortex memory. *Applied Physics Letters* **96**, 132506 (2010).

34      Vargas, N. M. *et al.* Asymmetric magnetic dots: A way to control magnetic properties. *Journal of Applied Physics* **109**, 073907 (2011).

35      Sanz-Hernández, D. *et al.* Artificial Double-Helix for Geometrical Control of Magnetic Chirality. *ACS Nano* **14**, 8084 (2020).

36      Parkin, S. S. P. & Mauri, D. Spin engineering: Direct determination of the Ruderman-Kittel-Kasuya-Yosida far-field range function in ruthenium. *Physical Review B* **44**, 7131 (1991).

37      Lapa, P. N. *et al.* Spin valve with non-collinear magnetization configuration imprinted by a static magnetic field. *AIP Advances* **6**, 056107 (2016).

38      Jaafar, M. *et al.* Control of the chirality and polarity of magnetic vortices in triangular nanodots. *Physical Review B* **81**, 054439 (2010).

39      Navarro, H. *et al.* A hybrid optoelectronic Mott insulator. *Applied Physics Letters* **118**, 141901 (2021).

40      Altbir, D., Kiwi, M., Ramírez, R. & Schuller, I. K. Dipolar interaction and its interplay with interface roughness. *Journal of Magnetism and Magnetic Materials* **149**, L246 (1995).

41      Vansteenkiste, A. *et al.* The design and verification of MuMax3. *AIP Advances* **4**, 107133 (2014).

42      Novosad, V. *et al.* in *NSTI* Vol. 1   308-311 (TechConnect Briefs, Anaheim, CA, 2005).

43      Novosad, V. *et al.* Nucleation and annihilation of magnetic vortices in sub-micron permalloy dots. *IEEE Transactions on Magnetics* **37**, 2088 (2001).

44      Morales, R. *et al.* Magnetization depth dependence in exchange biased thin films. *Applied Physics Letters* **89**, 072504 (2006).

45      Morales, R. *et al.* Role of the Antiferromagnetic Bulk Spin Structure on Exchange Bias. *Physical Review Letters* **102**, 097201 (2009).

46      Jaafar, M. *et al.* Field induced vortex dynamics in magnetic Ni nanotriangles. *Nanotechnology* **19**, 285717 (2008).

47      Bloemen, P. J. H., van Kesteren, H. W., Swagten, H. J. M. & de Jonge, W. J. M. Oscillatory interlayer exchange coupling in Co/Ru multilayers and bilayers. *Physical Review B* **50**, 13505 (1994).

48      Vavassori, P. *et al.* Magnetostatic and exchange coupling in the magnetization reversal of trilayer nanodots. *Journal of Physics D: Applied Physics* **41**, 134014 (2008).

49      Parkin, S. S. P., Bhadra, R. & Roche, K. P. Oscillatory magnetic exchange coupling through thin copper layers. *Physical Review Letters* **66**, 2152 (1991).

50      Ay, F. *et al.* Magnetic Properties of Fe/Ni and Fe/Co Multilayer Thin Films. *Applied Magnetic Resonance* **48**, 85 (2017).

51      Gliga, S., Yan, M., Hertel, R. & Schneider, C. M. Ultrafast dynamics of a magnetic antivortex: Micromagnetic simulations. *Physical Review B* **77**, 060404 (2008).

52      Gupta, S. *et al.* Important role of magnetization precession angle measurement in inverse spin Hall effect induced by spin pumping. *Applied Physics Letters* **110**, 022404 (2017).

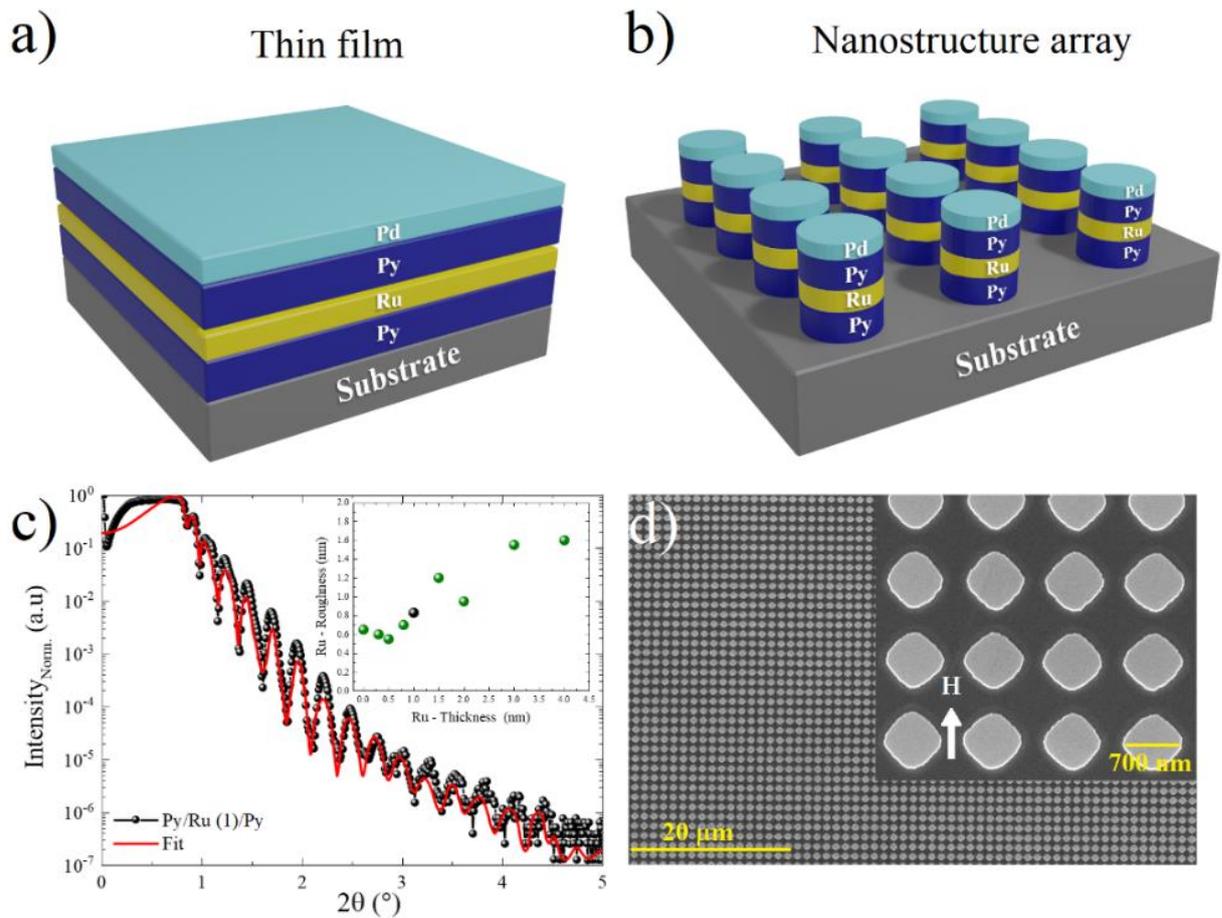

**Figure 1.** Schematics of the trilayer structure and the dot array. a) General structure of the trilayer, b) single cell of the nanostructure array, c) Representative example of an X-ray reflectivity of the trilayer in a continuous reference sample used to characterize layer thicknesses, (inset) Ru roughness evolution with spacer thickness d) Scanning Electron Microscopy image of the diamond-shaped nanostructure array. The arrow indicates the direction of the H field.

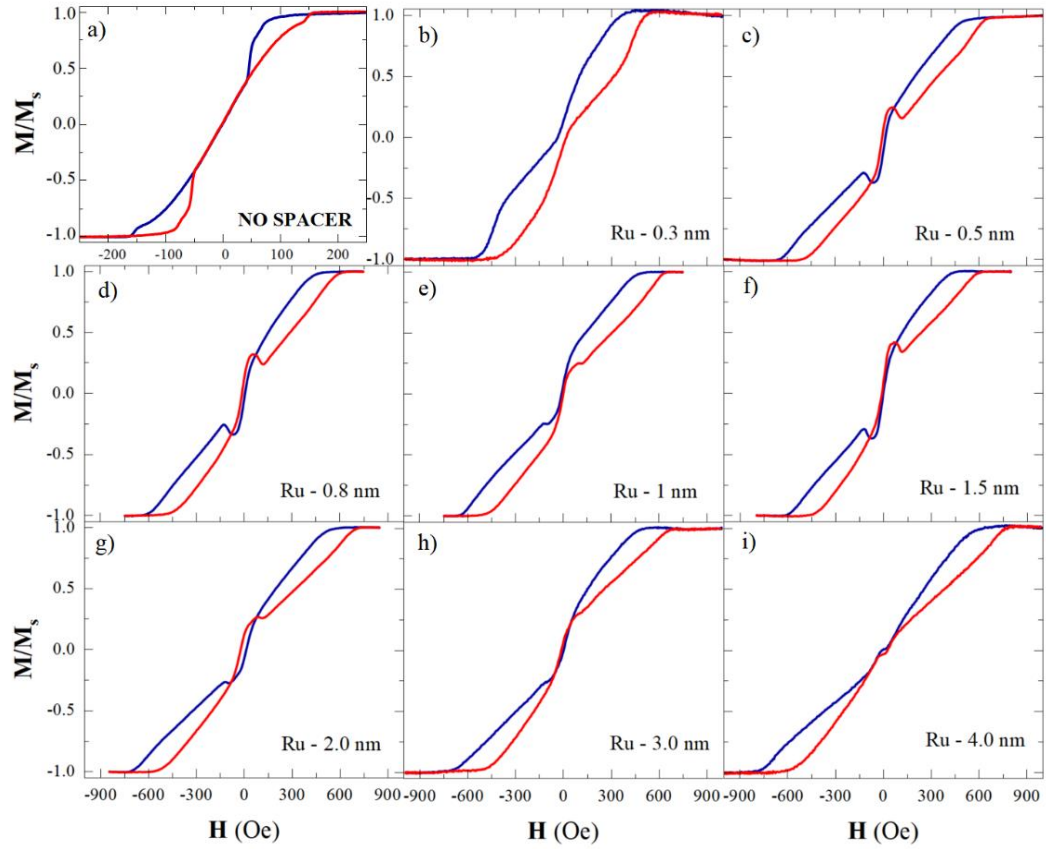

**Figure 2.** Longitudinal MOKE hysteresis loops of the Py(60) and Py(30)/Ru($t_{Ru}$)/Py(30) [$t_{Ru}$ =0.3, 0.5, 0.8, 1, 1.5, 2, 3, 4 nm] arrays of 700 nm nanostructures with the magnetic field parallel to the main axis of the square array in the substrate plane as shown in the inset of Figure 1d). The blue and red curves correspond to decreasing and increasing field directions, respectively.

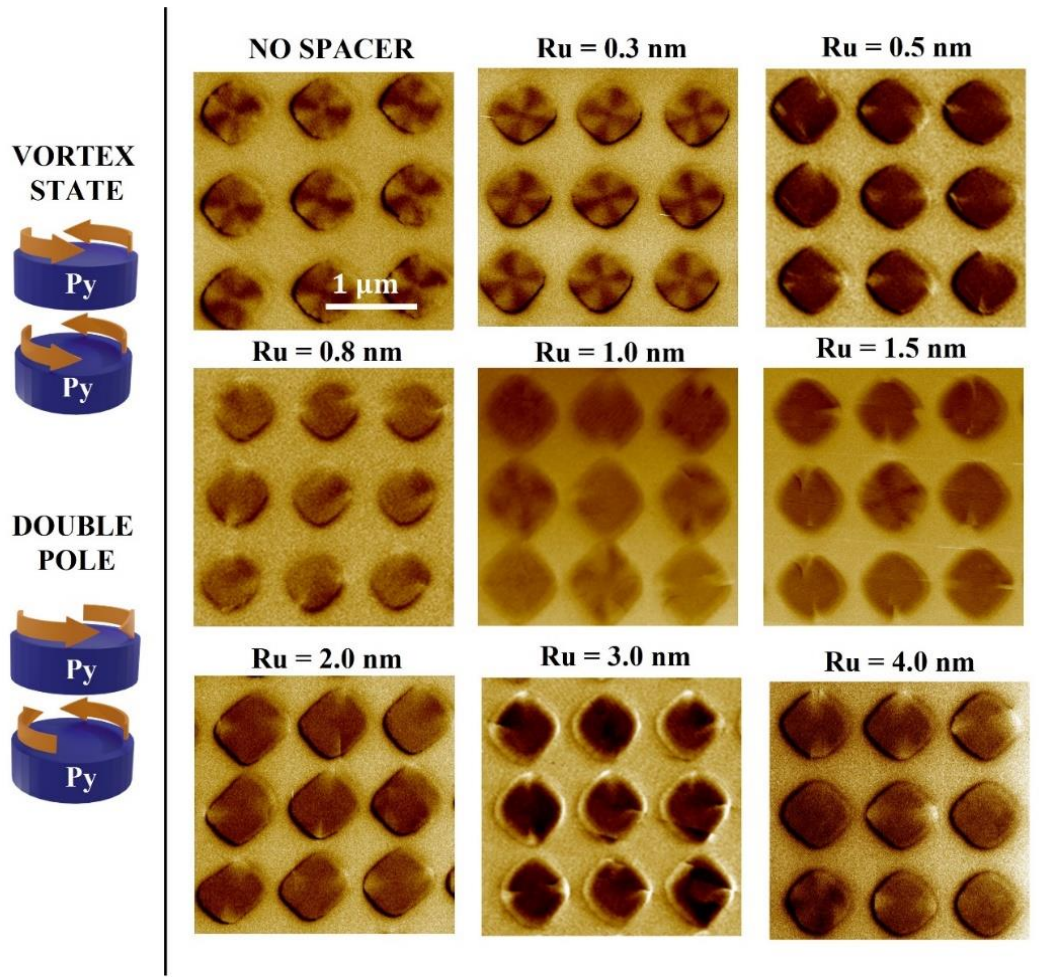

**Figure 3.** Magnetic Force Microscopy images of the magnetic textured formed in the single Py (60 nm) and in Py(30 nm)/Ru ($t_{Ru}$)/Py(30 nm) trilayers nanostructure arrays with increasing Ru spacer thickness (indicated above each image). The cartoon in the left panel of the figure schematizes the relative magnetization between both layers in the vortex state and in the double pole state.

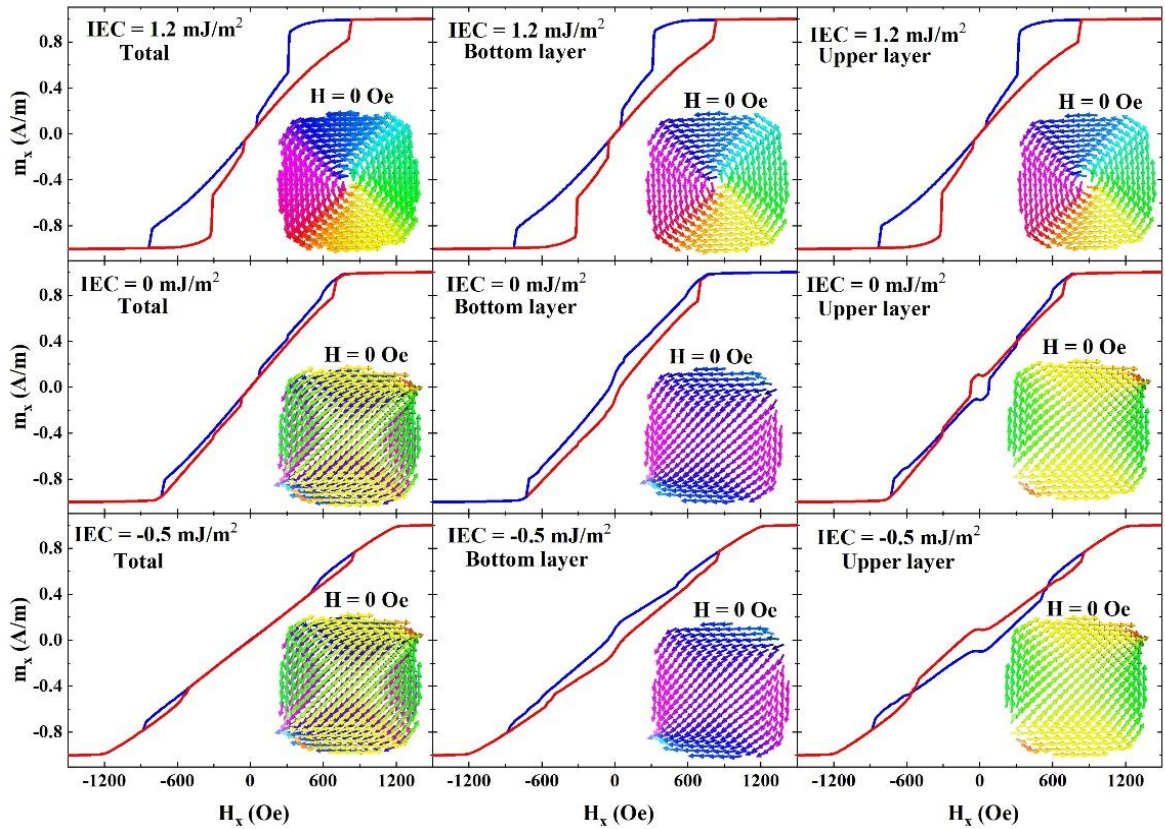

**Figure 4.** Micromagnetic simulations: magnetization map at remanence and the hysteresis loops corresponding to each bottom and upper Py layer and the total magnetization[36,47]. The blue curves are scans from positive to negative, while the red ones are from negative to positive fields. The color code on the magnetization map indicates the direction of in-plane magnetization: the same color corresponds to regions with equal magnetization direction. On this color scale, yellow/green and blue/magenta are perpendicular to each other; yellow/blue and green/magenta are antiparallel. In the upper map, the magnetization rotates counterclockwise and the transition between equally magnetized regions is shown in light colors.

| Ru thickness (nm) | MFM vortex | MOKE crossed cycle | Sign of $J_{ex}$ [36,47] |
|---|---|---|---|
| **0.3** | YES | NO | NEGATIVE - FM |
| **0.5** | NO | YES | POSITIVE - AFM |
| **0.8** | NO | YES | POSITIVE - AFM |
| **1.0** | YES | NO | NEGATIVE - FM |
| **1.5** | NO | YES | POSITIVE - AFM |
| **2.0** | NO | YES | POSITIVE - AFM |
| **3.0** | NO | YES | POSITIVE - AFM |
| **4.0** | Random | NO | NEGATIVE – FM |

**Table I.** Experimental results and sign of exchange coupling for various Ru thicknesses.